# Toward a Science of Autonomy for Physical Systems: Healthcare

Gregory Hager
hager@cs.jhu.edu
Johns Hopkins University

Eric Horvitz
horvitz@microsoft.com
Microsoft

In Star Wars Episode V, we see Luke Skywalker being repaired by a surgical robot. In the context of the movie, this doesn't seem surprising or disturbing. After all, it is a long, long time ago, in a galaxy far, far away.  It would never happen here. Or could it?  Would we accept a robot as our doctor, our surgeon, or our in-home care specialist? Imagine walking into an operating room and no one was there. You are instructed to lie down on the operating table, and the OR system takes over. Would you feel comfortable with this possible future world?

While we're happy to entertain the idea that autonomous systems will grow our food, build our buildings, and drive us to work, somehow the idea that autonomous physical agents may one day provide key healthcare services seems harder to envision or accept. One might argue that it is our relative unfamiliarity with healthcare that creates this feeling.  We drive our car every day and have an understanding of how cars work, but we never take out an appendix.  Perhaps the realities of medical interventions are just too far from our experience to understand how autonomous robotic systems might enhance our healthcare. Two aspects of healthcare may make it hard to imagine robotic systems playing an important role in medicine:

- Healthcare is personal: Robots and computers are good at doing the same thing over and over again, but everyone is different. How could an autonomous system ever be trusted to do the right thing *for me*?

- Healthcare is social: Medicine is seated deeply in social constructs; it is people taking care of people. It is not just about data and diagnosis, but about understanding the whole person and responding to their needs.

---

It is not hard to see that most people's response to the autonomous surgical suite is probably some combination of both elements. We want the comfort of knowing we are being treated as a person by a person.

So, where do these reflections take us? They suggest that a key goal in developing autonomous systems in healthcare should be to use autonomy to *enhance the human experience.* Autonomous systems can enhance the quality of human-human engagement by reducing repetitive and unrewarding activities, by making the use of people when they are most effective, and by making the healthcare organization a more people-friendly place. In what follows, we describe several ways that autonomy can enhance the quality, effectiveness, and cost of healthcare—and we note where some of these innovations are already underway[2]

## Healthcare as a Service Industry

The goal of healthcare is to serve patients by providing high quality and effective care. Where could autonomous systems enhance that service? Here is an example.

Imagine that you arrive at a typical healthcare facility for a diagnostic review to address some concerns about recent abdominal pain you are experiencing. Given your background, symptoms and your health factors, it is likely that you are experiencing obstruction of the cystic duct. After some tests on subsequent visits, it turns out that a specialist finds that there's a high likelihood you'll need your gall bladder removed. Today, this procedure, called a cholesystectomy, is typically an out-patient procedure. You start the search for a surgeon and a time to schedule the surgery. Meanwhile, your abdomen continues to hurt and you wonder about the associated risk factors.

Now, imagine a time in the future when you are met at reception by an autonomous concierge, a mobile robot equipped with basic interaction, video-conferencing capabilities and biometric measurement technologies (blood pressure, heart rate, and so forth). The system performs an initial assessment of your condition through an interview and, based on likely diagnoses (including your gallbladder issue), it begins to schedule your visit. It starts by immediately contacting a specialist who speaks with you, confirms the likely diagnosis, and listens to your concerns. The concierge then leads you through a series of diagnostic tests (all dynamically scheduled so you don't have to wait), all the while providing you updates on what it expects your day to be, and answering your questions. After the tests, it takes you to meet the specialist who confirms you need surgery.

---

[2] http://www.informationweek.com/mobile/10-medical-robots-that-could-change-healthcare/d/d-id/1107696?

After the conference, you decide not to wait, so you're immediately taken to a pre-operative suite. Automated systems have brought all of the necessary medications and equipment to your bed. A nurse arrives to supervise autonomous placement of central lines and other surgical prep. He has time to chat and address any concerns; he is not running from bed to bed dealing with the minutia of surgical prep for each patient. Finally, you are wheeled into the OR. There is one person there—the surgeon—who manages the entire surgical suite based on earlier research, including efforts funded by DARPA on triage in defense settings[3] When you wake up, the concierge is there and alerts the surgeon and the post-op nursing staff. All is well, and once the concierge assesses you to be ready to leave, the post-op nurse conducts a short interview, agrees with the assessment, and the concierge leads you to your autonomous car which takes you home.

How does autonomy contribute in this situation? Here are a few examples:

1) You have a constant companion while you're in the hospital. You're never left wondering what is going on.

2) Your time is optimized; the system can immediately interface with the hospital information system and work out how to best chart your course through the healthcare system.

3) Your treatment is optimized. You get immediate and personalized attention, and you don't need to wait for appointments, go home and return, and so forth. Every step of your care brings all of your information to bear.

4) The healthcare workers time is used to best effect. They can depend on your autonomous companion and the autonomous hospital infrastructure to make best use of their time, giving them more opportunity to do what they do best.

5) The hospital system is more efficient. Its resources can be used in a way that maximizes the use of equipment, people, and other resources.

## Healthcare as a Complex Supply Chain

Behind the scenes of your visit, the story continues. Hospitals are complex organizations that have to move an immense variety of material, equipment, and people. Managing inventory throughout the institution, and supplying resources where they are needed and when they are needed is no less complicated than an Amazon warehouse – indeed it is quite possibly more complex. This creates enormous opportunities for autonomous systems, for example:

---

[3] [http://www.ncbi.nlm.nih.gov/pubmed/19222048

1) Inventory and maintenance of supplies: Patient rooms, procedure rooms, the ICU, patient visit rooms – there is an enormous list of materials that need to be readily available. Lack of material is not only an inconvenience, it can affect patient safety. For example, Pronovost, in his landmark work on central-line infections[4] noted that one barrier to complying with best-practices for sterility was the simple lack of material – gowns, gloves, soap – when it was needed. When a gown wasn't available, but time was pressing, it was as likely as not that the healthcare worker would focus on the proximal patient need, not on where to go find gowns.

2) Patient services: Depending on their reason for stay, patients will often need special meals, drugs, or other similar deliveries. There are already companies who are beginning to explore this area[5], but only for limited cases. Broad-based autonomy would allow a deeper and wider use of physical agents to interact with patients and caregivers, and to provide a larger basket of services.

3) Smart equipment allocation: Hospitals are a capital-intensive business. This is in part because it is more efficient to have redundant equipment that is able to remain at a fixed station, than to hire labor and create the organization to move equipment where it is needed. Not every bed needs a respirator, IV unit, and cardiac monitor all the time, but every bed might at some point. Creating equipment that is able to autonomously relocate itself as needed would lower the capital investment of the hospital, and contribute to lower capital costs.

In summary there are many opportunities to use of same ideas that are being pioneered in merchandising and manufacturing in the healthcare environment. But, unlike merchandising and manufacturing, these systems will need to operate in a highly dynamic, people-intensive and unstructured environment – properties that challenge today's approaches to autonomy.

## Healthcare as a Skill and Knowledge Industry

The amount of information that a healthcare worker brings to bear in their job is growing exponentially. There are some estimates that by 2020, a typical clinician may have hundreds of relevant factors (physiological data, lab tests, patient history, genomics and so forth) to weight in making a diagnosis. While there are clear needs for methods for automated diagnosis, this also introduces the opportunity to create physical autonomy that can augment treatment.

---

[4] http://www.nejm.org/doi/full/10.1056/NEJMoa061115
[5] http://www.computerworld.com/article/2877284/at-ucsf-medical-center-robot-aided-healthcare-is-here.html

One area of opportunity is in the automation of certain aspects of interventions. For example, the growing prevalence of navigated (position-measured relative to the body) tools for needle-based interventions sets the stage for creating fully autonomous execution of certain procedures – for example image-guided biopsy or ablation. Further, as autonomous systems perform procedures, they will create data (images, tool movements, etc.) that can be used to improve performance through retrospective analysis or learning. The value of automation will be proven through better access, more consistent outcomes, and better quantification of what has been done through direct measurement.

A second possibility lies in the observation of, and collaboration with, doctors or surgeons as they perform their work. For example, the Intuitive Surgical da Vinci robot is now used in well over 0.5 million procedures per year. Imagine capturing all of this data, and building a universal database of "best practices" for specific surgeries. By creating intelligent observers who can then monitor a surgery, it would be possible to provide a basis for online feedback, mentoring, and learning for human surgeons, with the effect of improving both training and patient safety and outcomes. These observers can also provide guidance for surgeons who are performing a procedure they have not recently performed, and provide enhanced system performance by subtle modification of the surgical robot and OR systems for any point in the procedure.

A final value of autonomy is to create systems that enable doctors and surgeons to have "super-human" performance. For example, autonomous systems that can find and inject into a microscopic blood vessel in the eye, or which can detect and remove pre-cancerous cellular clusters invisible to the naked eye, or travel through and arterial system to treat disease using the bodies own natural highways.

Many of these ideas are already emerging in research laboratories around the world. Just as factory automation or semi-automated driving has created a platform for autonomous factories and autonomous driving, so to can we expect healthcare to experience similar transformations.

## Promise of Autonomous Systems to Shift Care to the Home

Beyond the value of autonomous robotics in hospitals, autonomous and semi-autonomous robotic systems promise to help with home-based healthcare. Polls have found that families and patients would prefer home-based care if possible. A Harris poll in 2010[6] found that over 75% of Americans would select home-based health care loved ones' instead of shifting to nursing facilities and other care facilities. Results also showed that a majority of families would prefer having terminally ill family members cared for at home with the help of professional health aide where necessary. The survey also found that over half of people polled would prefer that an elderly family member who is recuperating from surgery would prefer home health care over any other facility. An earlier Harris survey on end-of-life care in 2002 found that over 85% of Americans believe that people with terminal illness would most like to receive end-of-life care at home. However, over 70% of deaths in America today happen in professional healthcare facilities.

---

[6] http://www.prnewswire.com/news-releases/majority-of-americans-agree-theres-no-place-like-home-for-care-of-elderly-family-members-106949588.html

Autonomous and mixed-initiative systems can enable a shift of care of loved ones from expensive, impersonal sterile clinical facilities to home-based care. They can help to address this gap the reality of end-stage care and the preferences of families and of patients. The gap has been partly attributed to concerns for home-based care for a severely ill person. Taking care of an ill loved one can be overwhelming for families.

Beyond elderly and terminally ill patients, care can be shifted to the home for other conditions that require ongoing monitoring and interventions. Currently, such services are in chronically short supply, and the cost of such services is an immense load on patients, their families, and the healthcare system. New kinds of autonomous systems can help to move care to the home for patients with chronic conditions, patients undergoing rehabilitations, or patients waiting for a procedure (e.g. for a heart or lever transplant) to the home.

Imagine extending the concierge in the opening section to accompany the patient to their home. There, such systems can continue to provide monitoring, contact with primary care physicians and specialists, and the patient's support network as long and as often as needed.

Autonomous systems can be developed for addressing specific challenging problems that make home care difficult. For example, consider the challenge of maintaining the airway of an ill or incapacitated person. In many conditions, airways can be compromised by the failure of a patient to clear mucosal secretions on their own, based on the lack an ability to cough and swallow in a normal manner. These patients can recurrently face trouble with breathing and also endure an ongoing risk of acquiring a serious pulmonary infection. Challenges with airway management are ongoing threats to patients on chronic ventilation systems and to those with tracheostomies. However, other patients who are frail and incapacitated face airway challenges, including elderly people who have been weakened by recent surgery, people facing neuromuscular disorders like ALS, and people in the end stages of metastatic cancer.

Developing an autonomous system for suctioning patients at the right times and in a gentle, careful, and effective manner is a representative high-stakes grand challenge for robotic autonomy home-based healthcare. Such an autonomous system should be endowed with competencies that enable it to continually listen and watch for breathing difficulties, to have the ability to establish communications and to engage with patients via a grounding on the patient's goals and needs (even if just via understanding and responding to the patient's facial gestures, movement of the eyes, or subtle nods) and then to work with the physicality and the physics of suctioning in a gentle and careful manner, watching out for the comfort of the patient.

The *autonomous airway management challenge* is representative of numerous other challenges that can be tackled by the robotics community for addressing key problems and pain points in healthcare, with applications to enhancing home-based, as well as hospital-based care. In summary, there is an urgent need to aim advances in robotics at providing vigilant, effective, but gentle and caring autonomous care-providing systems. Such systems will have influences on reducing the costs of healthcare while raising the quality of the lives of patients and their families.

## Final Thoughts

These are but a few ideas for how physical autonomy might impact the healthcare space. Looking across these applications, we can see there are several exciting and challenging research directions that would enable future progress in this area. These include:

1) Robust navigation is complex and dynamic environments, and appropriate action (or lack thereof) in response to surrounding events.

2) Human-robot interaction and task performance that is sensitive to a wide range of human physical and mental capabilities and social situations.

3) The development of flexible manipulation capabilities that can work with small objects (pills, vials, needles), flexible objects (gowns, gloves, suture), and delicate objects (glass, instruments, and patients)

4) Methods to model and understand complex manipulation tasks, and which are able to quickly relate online data into stored representations or records – for example being able to model a surgery, and to determine what the surgeon is doing as a procedure progresses.

Major obstacles for work on autonomous robotic systems in healthcare include the special attributes of medicine as strongly people-centric, high-stakes, safety-centric, and privacy conscious. A final set of challenges stem from the need to create realistic "living labs" and associated data sets that allow testing and experimentation of new ideas and systems, and which form the translational conduit that will finally make Luke Skywalker's "in-silico" clinician a reality.

*For citation use*: Hager G. D. & Horvitz E. (2015). *Toward a Science of Autonomy for Physical Systems: Healthcare*: A white paper prepared for the Computing Community Consortium committee of the Computing Research Association. http://cra.org/ccc/resources/ccc-led-whitepapers/

This material is based upon work supported by the National Science Foundation under Grant No. (1136993). Any opinions, findings, and conclusions or recommendations expressed in this material are those of the author(s) and do not necessarily reflect the views of the National Science Foundation.